\documentclass[%
 aip,
rsi,%
amsmath,amssymb,
 reprint,%
]{revtex4-1}

\usepackage{graphicx}
\usepackage{dcolumn}
\usepackage{bm}
\usepackage{multirow}
\usepackage{color}

\usepackage{tikz}

\newcommand\copyrighttext{%
  \footnotesize This article may be downloaded for personal use only. Any other use requires
prior permission of the author and AIP Publishing.
This article appeared in 
Review of Scientific Instruments 89, 093103 (2018)
and may be found at 
https://doi.org/10.1063/1.5039344}
\newcommand\copyrightnotice{%
\begin{tikzpicture}[remember picture,overlay]
\node[anchor=south,yshift=-5pt] at (current page.south) {\fbox{\parbox{\dimexpr\textwidth-\fboxsep-\fboxrule\relax}{\copyrighttext}}};
\end{tikzpicture}%
}

\makeatletter
\AtBeginDocument{\let\LS@rot\@undefined}
\makeatother

\begin{document}
\copyrightnotice
\preprint{AIP/123-QED}

\title[An open and flexible digital phase-locked loop for optical metrology]
{An open and flexible digital phase-locked loop for optical metrology} 

\author{Alex Tourigny-Plante}
\author{Vincent Michaud-Belleau}
\author{Nicolas Bourbeau H\'ebert}
\author{Hugo Bergeron}
\affiliation{%
Centre d`optique, photonique et laser, Universit\'e Laval, Qu\'ebec, QC, G1V 0A6, Canada
}%
\author{\\J\'er\^ome Genest}
\email{jgenest@gel.ulaval.ca}
\affiliation{%
Centre d`optique, photonique et laser, Universit\'e Laval, Qu\'ebec, QC, G1V 0A6, Canada
}%

\author{Jean-Daniel Desch\^enes}
 \email{octosigconsulting@gmail.com}
\affiliation{%
Centre d`optique, photonique et laser, Universit\'e Laval, Qu\'ebec, QC, G1V 0A6, Canada
}%
\affiliation{%
OctoSig Inc. Qu\'ebec, Canada
}%
\date{7 May 2018}

\begin{abstract}
This paper presents an open and flexible digital phase lock loop optimized for laser stabilization systems. It is implemented on a cheap and easily accessible FPGA-based digital electronics platform (Red Pitaya) running a customizable open-source firmware. A PC-based software interface allows controlling the platform and optimizing the loop parameters remotely. Several tools are included to allow measurement of quantities of interest smoothly and rapidly. To demonstrate the platform's capabilities, we built a fiber noise canceler over a $400$~m fiber link. Noise cancellation was achieved over a $30$~kHz bandwidth, a value limited mainly by the delays introduced by the actuator and by the round-trip propagation over the fiber link. We measured a total latency of $565$~ns for the platform itself, limiting the theoretically achievable control bandwidth to approximately 225 kHz.
\end{abstract}

\maketitle

\section{Introduction}

The use of digital electronics for optical systems and servo loops is steadily gaining traction. For example, digital phase meters \cite{WAN17} are used for gravitational waves detection \cite{SHA06, McRae:14,Isleif:14} and digital phase-locked loops (DPLL) are deployed for frequency comb locking \cite{SIN15, Lezius:16}, Doppler cancellation in optical links\cite{Olaya:16,DRO13,7384737} as well as optical time transfer\cite{DES16}. This is due to several advantages of digital servos such as software reconfigurability, easy replication and the possibility of having advanced diagnostic tools\cite{Calosso:14}.

In applications where low latency and large control bandwidth are important, field programmable gate arrays (FPGAs) appear as platforms of choice thanks to their intrinsically parallel and reconfigurable architecture. Digital platform can also directly track the phase of the signal instead of its trigonometric representation, which causes ambiguity (i.e. $\cos2\pi=\cos4\pi$). This wider tracking possibility allows using tools to characterize the external system even when the free running system has phase noise greater than the linear region of the trigonometric function. However, the complexities associated with designing and programming a mixed mode electronics board where fast analog signals are digitized, processed and re-generated in real time are however usually not in a typical laser metrologist's tool chest and the price of FPGA-based systems is often prohibitive. These constitute important entry barriers preventing deployment of powerful digital platforms to a wider set of optical applications.

In this paper, we present an open DPLL platform built around the Red Pitaya, a cheap and easily accessible FPGA-based board \cite{REDPIT}. The Red Pitaya can be purchased for a few hundred dollars, a fraction of what commercial analog lock boxes cost. The platform described here consists of a firmware installed on a Red Pitaya board as well as a Python-based PC graphical user interface~(GUI) that remotely connects to boards in order to monitor and optimize the servo loop performance. The firmware and software are open-source and can be downloaded, along with the installation instructions and user guide, using the link in the references\cite{DPLL}. The software should be cross-platform, but the installation instructions are only provided for Windows and MacOSX.

This work is based on the open-source design presented in \cite{SIN15,NIST_DPLL} now ported and adapted to Red Pitaya boards so that custom hardware is no longer required. An optional internal voltage-controlled oscillator (VCO) has also been added to facilitate the integration with common optical components, for example the acousto-optic modulator (AOM) used to frequency-shift an optical field in a fiber noise canceled link, as presented in section \ref{sec:FiberNoiseCanceler}. Although this platform is not limited to optical metrology, the tools described in section \ref{sec:firmwareArchitecture} were optimized with that application in mind since the residual phase is not as important when dealing with much higher optical carrier frequencies (compared to RF signals)  and since there is a stronger focus on lowering the cycle slip probability.

For some applications, FPGA platforms can be highly optimized to perform a single task\cite{Olaya:16} while in other cases the goal is to offer as many digital instruments as possible \cite{Moku:Lab,PyRPL}. The platform presented here sits in between these two extremes. The goal is to provide laser scientists with a straightforward access to a fast and flexible DPLL. With the design complexity tackled on generic off-the-shelf commercial electronics, the replication cost of this system is very low and the DPLL can easily be deployed in a variety of optical applications. 

\section{Firmware architecture}
\label{sec:firmwareArchitecture}

Red Pitaya boards are based on a Xilinx Zynq~7010 FPGA and are programmed using Vivado 2015.4 with a mix of VHDL and Verilog. The board is equipped with two analog inputs and two analog outputs. The analog-to-digital and digital-to-analog converters (ADCs and DACs) have a resolution of $14$~bits (a $10$~bit version is also available and is compatible with the DPLL described here) and are, by default, driven by the same $125$~MHz crystal that is clocking the FPGA. The Red Pitaya's inputs are equipped with 50~MHz low-pass anti-aliasing filters \cite{REDPIT_docs}. Each board can thus provide two DPLL that can be independent or interlinked, as described below. 

\begin{figure*}[htbp]
  \includegraphics[scale = 0.93]{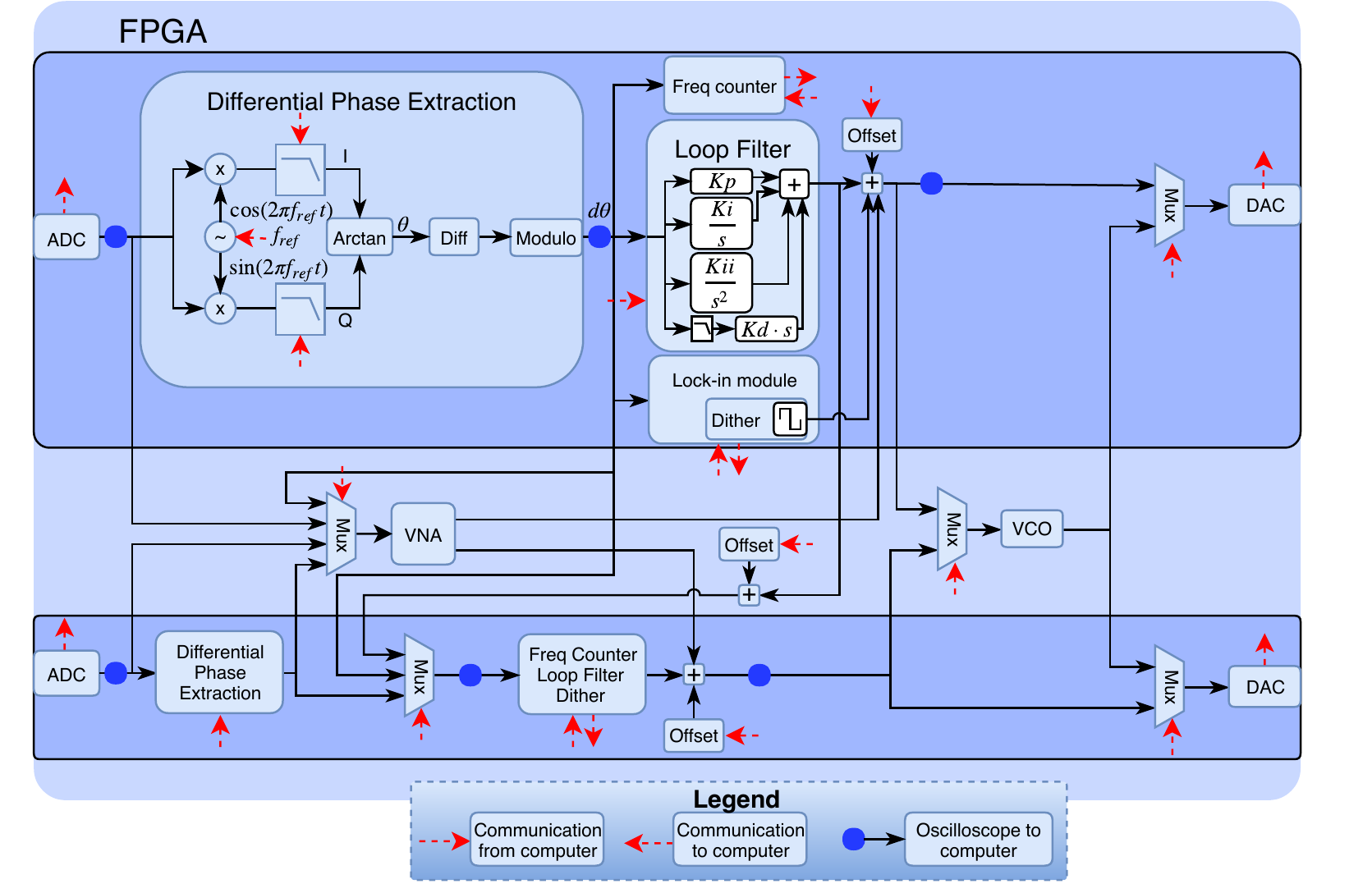}
  \caption{\small Complete block diagram of the DPLL platform's firmware. Red arrows represent the communication with the PC software and blue dots represent the test points at which signals can be shown in the PC software. The second channel is a duplicate of the first. ADC: Analog-to-digital converter, MUX: Multiplexer, VNA: Vector network analyzer, VCO: Voltage-controlled oscillator, DAC: Digital-to-analog converter.} 
\label{fig:bloc_diag_complete}
\end{figure*}

Figure \ref{fig:bloc_diag_complete} shows the complete representation of the DPLL architecture. For each channel, the signal is acquired by an ADC. Its in-phase and quadrature (I/Q) components are extracted via multiplication with the sine and the cosine of a reference signal whose frequency is defined by a rational fraction on 48 bits chosen by the user ($f_{ref} = k/2^{48}\cdot f_{clk}$) \cite{webb1978digital}. The I/Q signals are low-pass filtered digitally with a selectable bandwidth (3.75~MHz, 15.5~MHz or 31~MHz). An arctangent operation is then used to extract the phase error and a difference of successive phase measurements (numerical derivative) is used to alleviate dynamic range issues with the numerical representations in the FPGA since a phase ramp is unbounded while its derivative is bounded. The modulo $[-\pi,\pi]$ operator handles the wrapping of the phase slope produced by the arctan. The phase increment signal is the input of the loop filter implementing a proportional/integral/double integral/derivative ($\textrm{PII}^2\textrm{D}$) controller \cite{ang2005pid}. The double integrator term ensures a null static error even with the numerical phase derivative previously made. The resources usage for this platform is presented in table \ref{tab:utilization}. 

\begin{table}[htbp]
\begin{tabular} {l||c|c}
Resource & Used & Available \\ \hline
 Look-up table (LUT) & 8449 & 17600 \\
 LUT RAM & 370 & 6000 \\
 Flip-Flops (FF) & ~12505~ & 35200 \\
 Block RAM (BRAM) & 53.50 & 60 \\
 DSP Slices & 69 & 80 \\
 Global clock buffer (BUFG) & 9 & 32 \\
  Mixed-mode clock
manager & \multirow{2}{*}{1} & \multirow{2}{*}{2} \\
(MMCM) & & \\
 Phase-locked loop (PLL) & 2 & 2
\end{tabular}
\caption{Red Pitaya resources usage.}
\label{tab:utilization}
\end{table}

The ADC chips used in the Red Pitaya have a nominal time-domain signal to noise ratio (SNR$_t$) of $73$~dB (compared to a theoretical $86$~dB for $14$~bits) but the overall performance of the Red Pitaya system with $14$~bits ADCs is a SNR$_t$ of $63$~dB, corresponding to 10.2 effective number of bits (ENOB). The time domain SNR$_t$ is equivalent to the spectral SNR$_f$ in a bandwidth from 0 to $f_s$/2. The spectral SNR$_f$ is the SNR seen with a bandwidth of $BW$~Hz and is described by \mbox{$\textrm{SNR}_f=\textrm{SNR}_t+10\log_{10}(\frac{f_s}{2BW})$}. For instance, when operating at a $125$~MS/s sampling rate, this provides a SNR over $72$~dB in a $7.5$~MHz bandwidth, which is well above the SNR usually obtained in beats involving frequency combs \cite{Deschenes:15}.

The output of the loop filter is encoded into 16 bits. When the VCO is not connected, the $14$ most significant bits (MSBs) are directly sent to the $14$-bit DAC and mapped to a voltage between $-1$ and $+1$~V at the SMA outputs. When the VCO is connected, the $16$ bits are instead used to represent a frequency between $0$ and the Nyquist frequency ($f_s/2$). The 16 bits are mapped such that $0 = 0~\textrm{Hz}$ and $2^{16}-1 = 62.5~\textrm{MHz}$, corresponding to a gain of 31.25~MHz/V. Adding an offset at the output via the control software allows adjusting the quiescent frequency. The desired tone is generated using an internal direct digital synthesis (DDS) and this signal is sent to the DAC with a user selected amplitude and DC offset.

The addition of a dither signal allows for an easy characterization of the system to be controlled. The dither module generates a square wave with a selectable amplitude and frequency. This signal is added just after the loop filter's output and thus it constantly excites the system being controlled. Lock-in detection \cite{doi:10.1119/1.17629} of this square wave at the input side of the module allows measuring the controlled system's gain in Hz/V, including its sign. This gives the operator a constantly updated knowledge of the loop's sign and is useful for performing loop tuning in relevant units.

The loop filter parameters are user selectable in the GUI. The proportional gain ($k_p$) is defined in dB relative to the open-loop DC gain of the system ($k_c$), such that the linear gain is $K_p = 10^{k_p/20}/{k_c}$. The integrator, double integrator and derivative are specified by their crossover frequencies, $f_i$, $f_{ii}$ and $f_d$, respectively. The integrator and the differentiator crossover frequencies can be defined either by the 0~dB crossover or by the $kp$ crossover. The double integrator crossover frequency is defined by the intersection with the integrator. Finally, a first-order filter implements a roll-off frequency for the differentiator ($f_{df}$). Table \ref{tab:LoopFilterEquation} presents the equations linking the loop filter gain values ($K_p$, $K_i$, $K_{ii}$ and $K_d$) to the user settings and the system parameters. A factor $2\pi/f_s$ takes into account the fact that the numerical derivative and integrators are relative to the sampling frequency ($f_s$).

When including the previous numerical derivative, the PII$^2$D controller in fact acts as a PIDD$^2$. For most systems being controlled, the double differentiator term would not be used and is normally turned off. However, this can be useful for example with a system containing two dominants first order roll-offs. 

\renewcommand{\arraystretch}{2.1}
\begin{table}[htbp]
\begin{tabular} {c||c|c}
Variable &~Relative to 0~dB~&~Relative to $K_p$~\\ \hline
$K_p$ & \multicolumn{2}{c}{\scalebox{1.2}{$10^{k_p/20}\frac{1}{k_c}$}} \\ \hline
$K_i$ & \scalebox{1.2}{$\frac{1}{k_c}f_i\frac{2\pi}{f_s}$} & \scalebox{1.2}{$K_p~f_i\frac{2\pi}{f_s}$}  \\ \hline
 $K_{ii}$ &  \multicolumn{2}{c}{\scalebox{1.2}{$K_i~ f_{ii}\frac{2\pi}{f_s}$}} \\ \hline
 $K_d$ & \scalebox{1.2}{$\frac{1}{k_c}\frac{1}{f_d}\frac{f_s}{2\pi}$} & \scalebox{1.2}{$K_p\frac{1}{f_d}\frac{f_s}{2\pi}$} \\ \hline
 D filter & \multicolumn{2}{c}{\scalebox{1.2}{$f_{df}\frac{2\pi}{f_s}$}} \\ \hline
\end{tabular}
\caption{Loop filter gain equations.  $k_c$: system open-loop DC gain, $f_s$: sampling frequency (125~MHz)}.
\label{tab:LoopFilterEquation}
\end{table}

Diagnostic information can be sent from the Red Pitaya to the PC-based control software. The spectrum of all signals represented by a blue dot in figure \ref{fig:bloc_diag_complete} can be computed and displayed in real-time in the GUI; temporal representations are also available. Baseband components are represented in a conventional I/Q diagram for analysis at-a-glance while the phase and frequency noise power spectral density (PSD) and total integrated phase noise are also continuously updated. A zero dead-time frequency counter with a $1$ second gate time allows monitoring and logging the lock's long-term performance. 

A vector network analyzer (VNA) integrated to the platform makes the measurement of the system's transfer function possible. Measurements of the magnitude and phase are displayed in the PC-based software, allowing for an easy tuning of the controllers, and data can also be exported for further analysis. When the internal VCO is used, its contribution is included in the measured VNA's transfer function. Thus, the measured frequency response includes the contributions of the VCO, external setup and delays associated to the FPGA processing with an input calibrated in Hz and an output calibrated in V.

Software controllable multiplexers allow using the FPGA in three different control scenarios. First, the platform can operate as two independent control loops each using their own ADC, demodulation block, loop filter and DAC. For instance, one channel can be used to lock the carrier offset of a frequency comb and the other to lock one comb tooth to a reference laser, as done by Sinclair \emph{et al.} \cite{SIN15}. A single Red Pitaya box can therefore fully reference a frequency comb. The only caveat in this mode is that only one VCO is available and thus only one of the channel can use it. 

The second control scenario consists in having two parallel loops using only one input. In that case, both channels share the same ADC and demodulation block. After the differential phase extraction, the error signal however proceeds independently in the two distinct loop filters with potentially different transfer functions. This allows controlling two different actuators acting on the same system with different frequency responses. For instance, one might want to close a loop with a fast but short-stroked piezoelectric actuator on one side and, on the other side, a slower actuator with longer stroke \cite{SIN15}. 

In the third scenario, it is possible to seed the output of the first channel to the input of the second filter loop. This configuration can be used to maintain the output of the first channel around a certain setpoint by controlling another actuator with the second channel. This method is used to stay in the active range of a fast actuator by relying on a slow actuator. For example, Sinclair \emph{et al.} used a slow piezoelectric actuator to stay within the dynamic range of a fast piezoelectric actuator\cite{SIN15}.

The whole system latency is $\tau= 407$~ns ($565$~ns with the VCO), which limits the noise rejection bandwidth to approximately $1/8\tau = 300$~kHz ($225$~kHz with the VCO). This estimation of the maximum bandwidth assumes that a loop is closed around a first order system with $\pi$/4 of phase margin. Of the $407$~ns latency, $207$~ns comes from the demodulation process and the remaining comes from the loop-filter, ADC and DAC.

The communication between the FPGA and the computer is made via a TCP connection controlled by the Python software. A notable feature of this updated version is the capability to disconnect the PC control software, allowing locks to operate in a standalone mode, and reconnect when supervision, data acquisition or modifications of the parameters are needed. 

\section{Fiber noise canceler}
\label{sec:FiberNoiseCanceler}
\begin{figure}[htbp]
\centering
\includegraphics{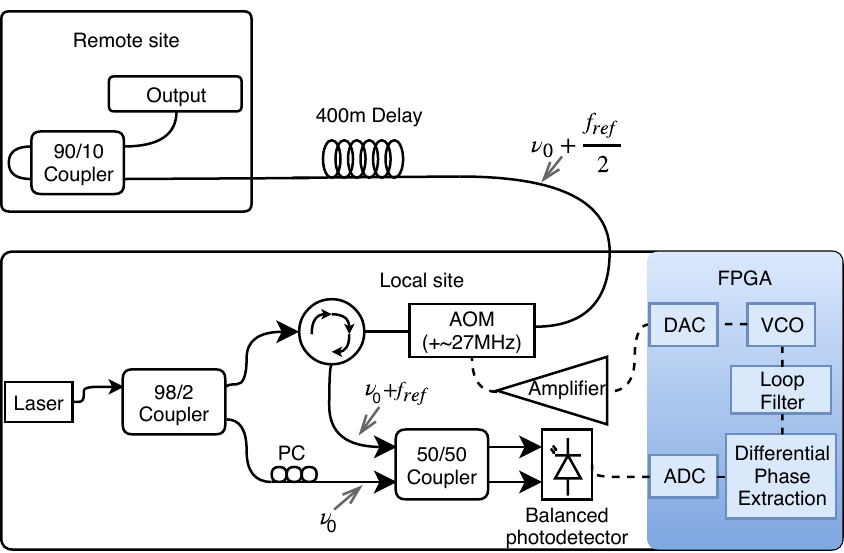}
\caption{\small Experimental setup for the fiber noise canceled link. Solid lines represent optical fibers and dashed lines represent electrical interconnects.  In our experimentation, $f_{ref}=54~MHz$. PC:~polarization controller, AOM:~acousto-optic modulator}
\label{fig:exp_setup}
\end{figure}

A fiber noise canceler (or Doppler canceler link) is a commonly used tool in optical frequency metrology \cite{Olaya:16,DRO13,7384737,Williams:08}. Here, we used a Doppler cancellation link to validate the operation of the DPLL on the Red Pitaya platform and to demonstrate new features, such as the integrated VCO. Figure \ref{fig:exp_setup} shows the experimental setup used to achieve the stable transfer of a laser frequency to a remote location over a fiber link. In the chosen configuration, part of the laser light reaching the remote site is reflected to the transmission site. This round-trip light thus experiences, at low frequencies, double the delay and double the phase fluctuations imposed by the link on the transmitted signal, as shown by Williams \emph{et al.} \cite{Williams:08}. Compensation for the measured fluctuations is done using an AOM fed by the Red Pitaya's new feature, the VCO, which is controlled by the DPLL.

At the local site, the beat note between the local oscillator (LO) and the reflected signal is fed to the DPLL. Its phase contains information about the fluctuations in the link's optical length that have to be corrected. As the nominal frequency of the AOM used here is 27~MHz, I/Q demodulation is performed with a reference frequency of 54~MHz in order to extract the phase error. The output offset is adjusted to set the quiescent frequency at $27$~MHz. 

\begin{figure}[t]
\center	\includegraphics{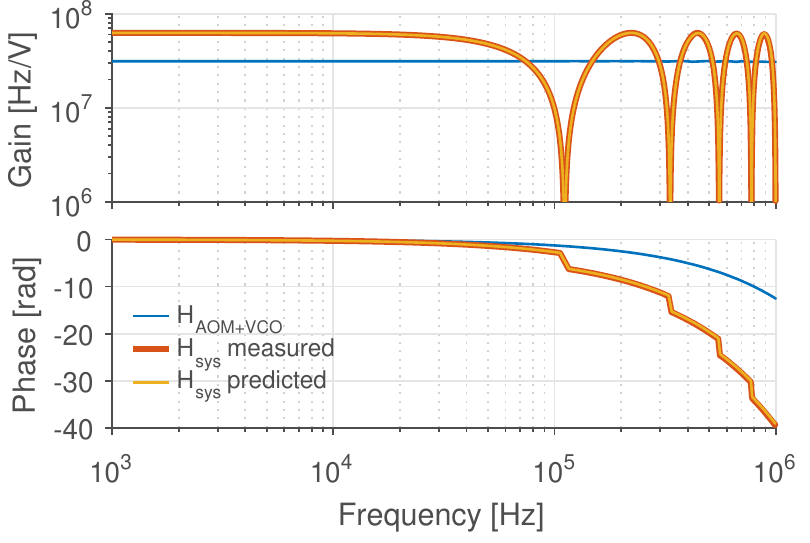}
\caption{\small Transfer function of the AOM-VCO combination (blue) and the complete Doppler cancellation link (orange), as measured by the platform's VNA. The AOM-VCO combination transfer function (blue) is use to determine the complete theoretical transfer function (red). The complete experimental transfer function (orange) matches the theoretical one.}
\label{fig:bode_complete}
\end{figure}

In order to characterize the AOM used in the experiment, we measured its transfer function with a different setup in which a beat between the input and output signals of the AOM was acquired. The resulting curve, displayed in blue in figure \ref{fig:bode_complete}, represents the transfer function of the combination of AOM and VCO. The AOM's latency, $1.5~\mu s$, was computed from the transfer function based on the approximation that it mainly acts as a delay in the frequency range of interest. Therefore, this delay is larger than the intrinsic FPGA's latency and will always lead to a significantly lower noise rejection bandwidth.

Multiplying the AOM+VCO transfer function by the fiber optic delay and the cosine square shape of the double pass through the AOM leads to an excellent superposition with the transfer function measured by the VNA for the system shown in figure \ref{fig:exp_setup}. The first zero of the cosine at $0.125$~MHz is determined by the fiber link differential delay which is here $2.0~\mu s$ ($400$~m). With a total effective delay of $4.0~\mu s$ (optical fiber $2.0~\mu s$, AOM $1.5~\mu s$ and FPGA $0.5~\mu s$), the maximum achievable closed-loop bandwidth in this demonstration is thus $31.25$~kHz.

\begin{figure}[htbp]
\includegraphics{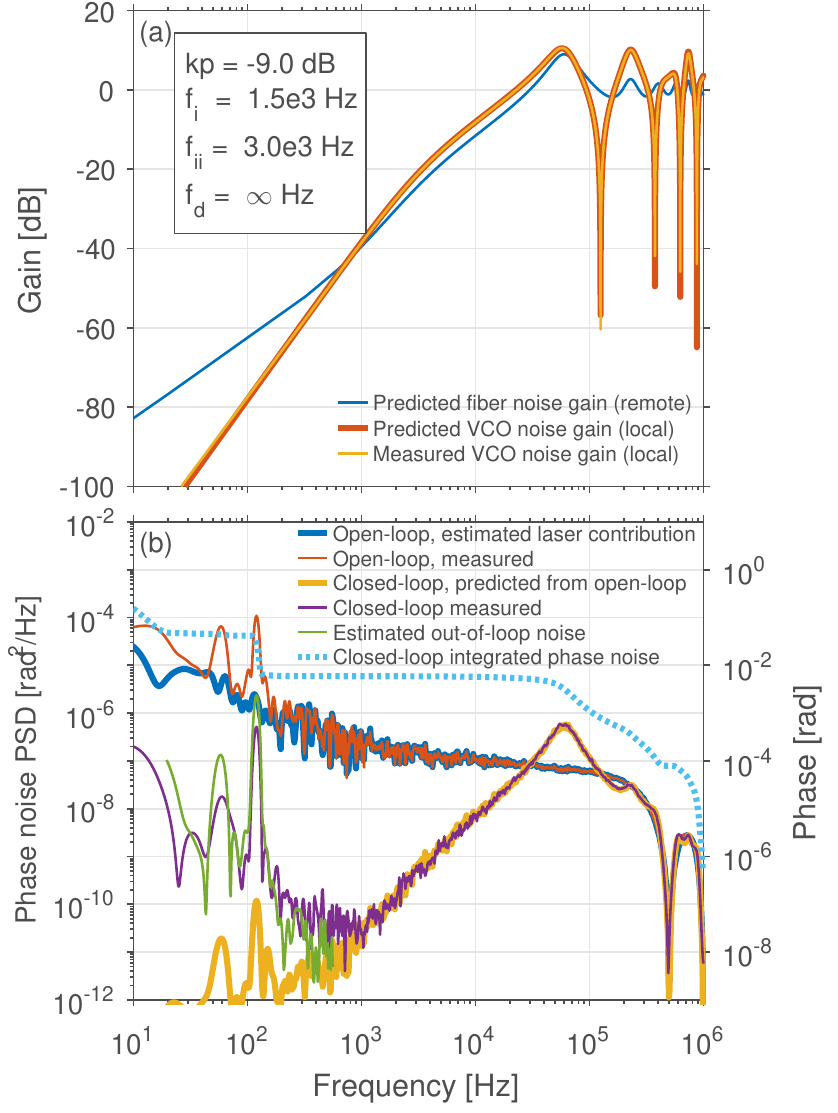}
\caption{\small \textbf{a)} Predicted closed-loop transfer functions for various noises and disturbances present in the system. The built-in VNA was used to measure the noise rejection transfer function for noise injected at the local end of the link (yellow). The remarkable agreement with the prediction (orange) yields credibility to other predictions based on the same set of estimated parameters. \textbf{b)} Comparison of the remote output phase noise PSD measured with an out-of-loop oscilloscope. The closed-loop prediction curve (yellow) is obtained by multiplying the open-loop phase noise PSD (orange) by the appropriate transfer functions, assuming that it is largely dominated by laser noise. A rough estimation of the out-of-loop noise (green) was performed by bypassing the 400 m delay to directly connect the AOM to the remote site. Obviously, this out-of-loop noise constitutes the main contribution to closed-loop noise below 1 kHz; since its level is lowered when we activate an anti-vibration table, we conclude that it is most likely due to residual building vibrations. The cumulative integrated phase noise (blue dots) is also shown.}
\label{fig:rejection-PSD}
\end{figure}

When the lock is active, the VNA can still be used to output perturbations in the system in order to characterize the closed-loop noise rejection, as shown in figure~\ref{fig:rejection-PSD}a). This figure displays the theoretical gain for uncorrelated distributed fiber noise at the remote link output, which is one of the noise our lock aims to cancel. To validate the model used to determine this transfer function, the experimental and predicted transfer functions for the VCO noise at the local site are shown. This experimental curve was easier to obtain since it could be measured without the addition of external components as the VCO was the only actuator. However, one could use a different channel to control an actuator to introduce disturbance at another point in the system. 

A visual representation of the obtained PSD is given in figure \ref{fig:rejection-PSD}b). This figure shows both closed-loop (purple curve) and open-loop (orange curve) PSDs at the remote site. In this figure, it is possible to observe that the system noise rejection bandwidth is around $30$~kHz, which fits with the expected maximum bandwidth. The PSDs presented here are the result of a beat between the laser and the remote signal. Therefore, the displayed interferometric phase noise is necessarily lower than the laser's phase noise below 1/(4$\tau$) \cite{0026-1394-53-5-1154}. We can thus conclude that, after propagation in the Doppler-canceled link, the phase noise added by out-of-loop fibers would be negligible with respect to laser phase noise.

\section{Conclusion}
In summary, we have demonstrated the use of an open-source DPLL platform by performing an optical frequency transfer over a fiber noise canceled link. The tools available on this platform allow to simplify the analysis of the system by offering an on-board VNA to measure transfer functions and by providing various real-time quantities such as PSD, I/Q, spectrum, time signal, frequency error and loop filter's transfer function. The different tools and control scenarios allow the use of this cheap and accessible platform for several optical applications. Since this platform is open source, we invite everyone to contribute to its improvement, for example by minimizing the latency to increase the achievable locking bandwidth.

\section{Acknowledgements}
The  authors  want  to  thank  the  support  from  the  Natural  Sciences  and  Engineering  Research  Council  of  Canada(NSERC).

\appendix
\nocite{*}
\bibliography{aipsamp}

\end{document}